\documentclass[12pt]{article}
\usepackage{graphicx,cite}
\usepackage{epsf,latexsym}
\usepackage{amssymb,amsmath}
\usepackage{dcolumn}% Align table columns on decimal point
\renewcommand{\baselinestretch}{1.5}

\begin{document}

%%%%%%%%%%%%---ARROWS FOR CONVERGENCE----%%%%%%%%%%%%%%%%%%%%%%%%%%%%%%%%%%%%%
\def\llra{\relbar\joinrel\longrightarrow}              %THIS IS LONG
\def\mapright#1{\smash{\mathop{\llra}\limits_{#1}}}    %ARROW ON LINE
\def\mapup#1{\smash{\mathop{\llra}\limits^{#1}}}     %CAN PUT SOMETHING OVER IT
\def\mapupdown#1#2{\smash{\mathop{\llra}\limits^{#1}_{#2}}} %over&under it%
%%%%%%%%%%%%%%%%%%%%%%%%%%%%%%%%%%%%%%%%%%%%%%%%%%%%%%%%%%%%%%%%%%%%%%%%%%%%%%

\begin{center}\Large{\bf New approach for alpha decay half-lives of superheavy nuclei and applicability of WKB approximation}
\end{center}

\begin{center}{\large Jianmin Dong$^{1,2,3,4,5,6}$, Wei Zuo$^{1,2,4}$,
   Werner Scheid $^{5}$ }\end{center}

%\maketitle
\begin{center}
\begin{enumerate}
\item
Research Center for Hadron and CSR Physics, Lanzhou University and Institute of Modern Physics of CAS, Lanzhou 730000, China
\item Institute of Modern
Physics, Chinese Academy of Sciences, Lanzhou 730000, China
\item Graduate
University of Chinese Academy of Sciences, Beijing 100049, China
\item School of Nuclear Science and Technology, Lanzhou
University, Lanzhou 730000, China
\item Institute
for Theoretical Physics, Justus-Liebig-University, D-35392 Giessen,
Germany
\item China Institute of Atomic
Energy, P. O. Box 275(18), Beijing 102413, People's Republic of
China
\end{enumerate}
\end{center}

%\maketitle

\begin{abstract}
\noindent The $\alpha$ decay half-lives of recently synthesized
superheavy nuclei (SHN) are calculated by applying a new approach
which estimates them with the help of their neighbors based on some
simple formulas. The estimated half-life values are in very good
agreement with the experimental ones, indicating the reliability of
the experimental observations and measurements to a large extent as
well as the predictive power of our approach. The second part of
this work is to test the applicability of the
Wentzel-Kramers-Brillouin (WKB) approximation for the quantum
mechanical tunneling probability. We calculated the accurate barrier
penetrability for alpha decay along with proton and cluster
radioactivity by numerically solving Schr\"odinger equation. The
calculated results are compared with those of the WKB method to find
that WKB approximation works well for the three physically
analogical decay modes.

\end{abstract}



%\maketitle
\section{Introduction}\label{intro}\noindent
Over the past decades, the syntheses of superheavy elements and
their lifetime measurement have been explored with a variety of
methods. The heavy elements with $Z = 107-112$ have been
successfully synthesized at GSI \cite{GSI}. Elements along with $Z =
113-116, 118$ have been produced at JINR-FLNR, Dubna \cite{YS0}.
Last year, two isotopes of a new element with atomic number $Z =
117$ were synthesized in the fusion reactions between $^{48}$Ca
projectiles and radioactive $^{249}$Bk target nuclei whose $\alpha$
chains terminated by spontaneous fission was observed in Dubna
\cite{117}, which fills the gap between the elements 116 and 118.
The element 114 was independently confirmed recently by the LBNL in
the USA \cite{LBNL1} and GSI \cite{GSI0}. A superheavy element
isotope $^{285}114$ was observed in LBNL last year \cite{LBNL2}, and
an isotope of $Z = 113$ has been identified at RIKEN, Japan
\cite{Jap}. Thus up to now superheavy elements with $Z = 104-118$
have been synthesized in experiment and consequently they offer the
possibility to study the heaviest known nuclei with greater detail.
However, their is no consensus among theorists with regard to what
should be the next doubly magic nucleus beyond $^{208}$Pb. Nearly
all of modern calculations predict the existence of a closed neutron
shell at $N = 184$. However, they differ in predicting the atomic
number of the closed proton shell. For instance, the
macroscopic-microscopic model predicts the shell gap at $Z = 114$
\cite{MM0,MM1,MM2}. The microscopic Skyrme-Hartree-Fock models give
$Z = 124, 126$ \cite{HFB1,HFB2,HFB3} and the relativistic mean-field
calculations suggest $Z = 120$ \cite{RMF1,RMF2,RMF3}. The magic
numbers $Z = 132$ and $N = 194$ were predicted from the
discontinuity of the volume integral at shell closures \cite{ZZ}. A
tremendous progress in experiments and the development of the
radioactive ion beam facilities have made it possible to reach the
island of superheavy elements.

The heaviest SHN decay primarily by the emission of
$\alpha$-particle terminated by spontaneous fission. Therefore, in
recent experiments, $\alpha$ decay has been indispensable for the
identification of new nuclides. Because the experimentalists have to
evaluate the values of the $\alpha$ decay half-lives, during the
experimental design, it is quite important and necessary to
investigate the $\alpha$ decay of SHN theoretically. Although
$\alpha$ decay is very useful for the study of the nuclei, a
quantitative description of them with a satisfying accuracy is
difficult. The $\alpha$ decay was firstly interpreted as a
consequence of quantum penetration of $\alpha$-particle by Gamow in
1928. At present, many theoretical approaches have been being used
to describe the $\alpha$ decay, such as the cluster model
\cite{Clu1,Clu2,Clu3}, the density-dependent M3Y (DDM3Y) effective
interaction \cite{DDM3Y1,DDM3Y2}, the generalized liquid drop model
(GLDM) \cite{GLDM1,GLDM2,GLDM22,GLDM,GLDM3}, the Coulomb and
proximity potential model \cite{CPPM}, the superasymmetric fission
model \cite{AS1,AS2}, the UMADAC method \cite{U1}, the coupled
channel approach \cite{coup1,coup2} and the universal curves for
$\alpha$ and cluster radioactivities in a fission theory
\cite{curv1}. Some physically plausible formulas also were employed
to calculate the $\alpha$ decay half-lives directly
\cite{AS1,JP,JPG,eq1,eq2,eq3}. Interestingly, the superasymmetric
fission theory for $\alpha$ and cluster decay has been extended by
some authors to study metallic cluster physics \cite{nano1,nano2},
which is an example of using the nuclear methods in nanophysics.

The half-life is extremely sensitive to the $\alpha$ decay $Q$ value
and an uncertainty of 1 MeV in $Q$ value corresponds to an
uncertainty of $\alpha$-decay half-life ranging from $10^{3}$ to
$10^{5}$ times in the heavy element region \cite{XXX}. In this work,
with the experimental $Q$ values, we carry out the half-life
calculations for the recently synthesized SHN by employing a
relationship between the $\alpha$ decay half-lives of neighboring
SHN that are established based on some simple semi-empirical
formulas. Differently from our approach, theoretical estimates for
the lifetimes by calculating the quantum mechanical tunneling
probability in a WKB framework is widely performed for the $\alpha$
decay along with other physically analogical decay processes. It is
pointed out that the WKB approximation works well at energies well
below the barrier height \cite{KH}. As a matter of fact, the
accuracy of the WKB approximation also depends on the shape of the
potential barrier as well as the decay energy. In this work, we
obtain the penetrability with a different method and show the
applicability of the WKB approximation in $\alpha$, proton and
cluster radioactivity.

This paper is organized as follows: In Section 2, a brief discussion
of the method and the calculated results along with the
corresponding discussions for the half-lives of SHN are presented.
The applicability of the WKB approximation for $\alpha$ decay,
proton and cluster emission are discussed in Section 3. Finally, a
brief summary is provided in Section 4.

\section{$\alpha$ decay half-lives of superheavy nuclei within a new approach}\label{model}\noindent
We start from Royer's \cite{Royer0} and Viola-Seaborg semi-empirical
(VSS) formulas \cite{VSS1,VSS2}. The Royer's formula is given by
\begin{equation}
\log
_{10}T(\text{s})=a+bA^{1/6}\sqrt{Z}+\frac{cZ}{\sqrt{Q}},\label{A}
\end{equation}
where $Q$ is in MeV and the parameter set varies for four types:
$a=-25.31$, $b=-1.1629$, $c=1.5864$ for even($Z$)-even($N$),
$a=-26.65$, $b=-1.0859$, $c=1.5848$ for even-odd, $a=-25.68$,
$b=-1.1423$, $c=1.592$ for odd-even and $a=-29.48$, $b=-1.113$,
$c=1.6971$ for odd-odd nuclei \cite{Royer0}. In our previous work,
this formula has been extended by taking into account the
centrifugal barrier to describe unfavored $\alpha$ decay. For
odd-mass nuclei, it is possible that some decays involve non-zero
$l$ values. However, as no experimental evidence is available for
the spin-parity of the levels involved in the decay, we have not
included the centrifugal barrier in the present calculations. The
VSS formula is given by
\begin{equation}
\log
_{10}T(\text{s})=(aZ+b)\frac{1}{\sqrt{Q}}+cZ+d+h_{\text{log}}.\label{B}
\end{equation}
Instead of using the original set of constants by Viola and Seaborg,
recent values $a=1.64062$, $b = -8.54399$, $c = -0.19430$,
$d=-33.9054$ being valid for the nuclei of four types are used
\cite{Dong}. $h_{\text{log}}$ accounts for the hindrances associated
with odd proton and odd neutron numbers but does not take an effect
in our calculations. Once the half-life of a nucleus $^{A_{0}}Z_{0}$
(reference nucleus) is known, the half-life of an other nucleus
$^{A}Z$ (target nucleus) with the same type can be derived. The
difference of the logarithms of half-life is written with Eq.
(\ref{A}) as
\begin{eqnarray}
S &=&\log _{10}T-\log _{10}T_{0} \nonumber\\
&=&b\left( A^{1/6}\sqrt{Z}-A_{0}^{1/6}\sqrt{Z_{0}}\right) +c\left( \frac{Z}{%
\sqrt{Q}}-\frac{Z_{0}}{\sqrt{Q_{0}}}\right),\label{C}
\end{eqnarray}
and with Eq. (\ref{B}) as
\begin{equation}
S=(aZ+b)\frac{1}{\sqrt{Q}}-(aZ_{0}+b)\frac{1}{\sqrt{Q_{0}}}+c(Z-Z_{0}).\label{D}
\end{equation}
Therefore, the half-life of the nuclei $^{A}Z$ can be obtained from
$T=10^{S}T_{0}$ with the help of its neighboring nucleus
$^{A_{0}}Z_{0}$. The two formulas can validate each other to obtain
more compelling results.

Now we focus on two simple cases. One is that the two nuclei are in
an isotope chain for which Eqs. (\ref{C}) and (\ref{D}) can be
further simplified, and the other is that the two nuclei belong to
an $\alpha$ decay chain. We estimated the $\alpha$ decay half-lives
of recently synthesized SHN $^{A}Z$ with the help of the reference
nuclei $^{A-2}Z$ and $^{A+2}Z$ by employing Eqs. (\ref{C}) and
(\ref{D}) without taking into account the uncertainty of the
experimental $Q$ values. The results are presented in Table 1
compared with experimental data. The results obtained with the DDM3Y
effective interaction and the GLDM are also shown for comparison.
The third column marks the experimental $\alpha$ decay half-lives,
and the columns 4 and 5 are the estimated ones from Eq. (\ref{C})
based on the Royer's formula and from Eq. (\ref{D}) based on the VSS
formula, respectively. The first half is obtained with the reference
nuclei $^{A-2}Z$ and the second half are obtained with the reference
nuclei $^{A+2}Z$. On an average, the DDM3Y results are slightly
larger than the experimental data while the GLDM values are lower
than the measured ones. In fact, the DDM3Y effective interaction and
GLDM are very successful because of the appropriate treatment on the
microscopic level in the DDM3Y calculation and the quasimolecular
shape in the GLDM in consideration of the difficulty in accurate
half-life calculation. Our calculated values are in very good
agreement with the experimental measurements which indicates our
method is a very effective approach to investigate the half-lives of
$\alpha$ decay when the experimental $Q$ values are given though it
is very simple in theoretical framework compared to the DDM3Y and
GLDM. The two approaches based on the Royer's formula and VSS
formula give nearly the same results implying the reliability of our
method to a certain extent. The half-life of $^{282}$113 is
underestimated by one order of magnitude with the present method and
the GLDM, which are possibly due to nonzero angular momentum
transfer or some nuclear structure effects such as the dramatic
deformation changes as suggested in Ref. \cite{shape} and the
influence of a possible neutron shell gap at $N=166$ on its daughter
nucleus. In Ref. \cite{gap}, it is suggested that $N=166$ is a
neutron shell gap in certain region within relativistic mean field
models. This nuclide warrants further experimental measurements with
higher statistics.

Apart from calculating the decay half-life, the present method is
also a useful approach to validate the experimental measurements.
The recently observed SHN still await confirmation by other
laboratories, which is not easy because the new SHN form an isolated
island that is not linked through $\alpha$ decay chain with known
nuclei. Therefore, the theoretical confirmations become important
and necessary. Since the half-lives of the reference nuclei are
taken from the experimental values, that the estimated results with
Eqs. (\ref{C}) and (\ref{D}) are in excellent agreement with the
experimental values suggests that the experimental half-lives are
themselves consistent with each other which confirms the reliability
of the experimental observations and measurements to a great extent.

In order to further confirm the above conclusions drawn from Table 1
and the predictive power of Eqs. (\ref{C}) and (\ref{D}), a
relationship between the $\alpha$ decay half-lives of the nuclei
belonging to an $\alpha$ decay chain is investigated here and the
estimated half-lives of SHN $^{A}Z$ are listed in Table 2. The
columns 4 and 5 are the estimated half-lives from Eqs. (\ref{C}) and
(\ref{D}) with the help of their daughter nuclei $^{A-4}(Z-2)$ while
the columns 6 and 7 are the estimated ones with the help of their
parent nuclei $^{A+4}(Z+2)$ respectively. The well agreement between
the evaluated values with the experimental ones further indicates
the predictive power of Eqs. (\ref{C}) and (\ref{D}) along with the
reliability of the experimental measurements. Of course, the
uncertainties in the measured values are large because of the
experimental difficulties and poor statistics. And our approach
underestimates the $\alpha$ decay half-life of $^{279}$111 as the
DDM3Y and GLDM. The reason for that is just the same as that for
$^{282}$113 mentioned above. Although the main shell effect has been
included in the $Q$ value to evaluate the half-life, the
preformation probability is also affected obviously by the shell
effect \cite{GLDM22}, which will lead to a large deviation of Eqs.
(\ref{A}) and (\ref{B}) together with Eqs. (\ref{C}) and (\ref{D})
for the nuclei around the magic numbers. However, as shown in Table
2, the half-lives of element 114 and the isotones $N=172$ are very
well reproduced by applying Eqs. (\ref{C}) and (\ref{D}). In other
words, it does not exhibit any evidence to show that $Z=114$ and
$N=172$ are shell closures in this region.

Finally, we predict the $\alpha$ decay half-lives of the nuclei
belonging to an $\alpha$ decay chain starting from $^{293}117$
without experimental values by employing Eqs. (\ref{C}) and
(\ref{D}). They are listed in Table 3 which may be useful for future
experimental measurements. The theoretical half-lives from Eq.
(\ref{C}) and Eq. (\ref{D}) agree very well with each other which
additionally confirms again the validity of our approach. The
deviation in the predicted values for $^{290}$115 may be large since
the experimental $Q$ value has a large uncertainty of 0.41 MeV. The
$Q$ value for this nuclide needs to be measured with a higher
accuracy. For other still unknown SHN, one can make predictions for
the $\alpha$ decay half-lives with the decay energies calculated
from the atomic mass evaluation of Audi {\it et al}. \cite{data} as
a substitute since the agreement with the experimental data on the
mass of the known heaviest elements is very satisfactory, or from a
formula for the $\alpha$-decay energy \cite{our}.

\section{Applicability of the WKB approximation }\noindent
We turn now to the applicability of the WKB approximation for
$\alpha$ decay. The interaction potential $V(r)$ is the sum of the
nuclear potential $V_{N}(r)$, Coulomb potential $V_{C}(r)$ and the
centrifugal barrier. We approximate the nuclear potential by
$V_{N}(r)=-A_{1}U_{0}/[1+\exp (\frac{r-R_{0}}{a})]$, with
$R_{0}=1.27A^{1/3}$ fm, $a=0.67$ fm, $U_{0}=\left[
53-33(N-Z)/A\right] $ MeV, $A_{1}$ the mass number of the emitted
particle, $N,Z,A$ the neutron, proton, and mass numbers of the
parent nucleus respectively. Here the single particle potential is
taken from Ref. \cite{NPZ}. The Coulomb potential is given by the
point-like plus uniformly charged sphere method with a parent
nucleus radius $R=1.28A^{1/3}-0.76+0.8A^{-1/3}$ fm \cite{ROYER}. As
an example, the barrier for $^{212}$Po$\rightarrow
^{208}$Pb$+\alpha$ is shown in Figure 1(a). Here only the barrier is
considered and we divide the barrier into a sequence of square
barriers, as shown in Figure 1(b). In principle, the barrier ranging
from $r_{1}$ to infinity should be taken into account for the
calculation yet it is unpractical. Therefore, we cut off the barrier
at a sufficiently large distance of $r_{2}=1000$ fm and the
potential barrier is divided into $n=60000$ parts with a step of
$h=(r_{2}-r_{1})/n$ without loss of accuracy. The wave function
$u(r)$ ($\Psi (\overrightarrow{r})=Y_{lm}(\theta ,\varphi )u(r)/r$)
of the emitted particle with $Q$ value in these $n$ regions can be
written as
\begin{eqnarray}
u_{1} &=&A_{1,1}\exp (ik_{1}r_{1})+A_{2,1}\exp (-ik_{1}r_{1}),  \nonumber \\
u_{2} &=&A_{1,2}\exp (ik_{2}r_{2})+A_{2,2}\exp (-ik_{2}r_{2}),  \nonumber \\
&&\vdots   \nonumber \\
u_{n-1} &=&A_{1,n-1}\exp (ik_{n-1}r_{n-1})+A_{2,n-1}\exp
(-ik_{n-1}r_{n-1}),
\nonumber \\
u_{n} &=&A_{1,n}\exp (ik_{n}r_{n})+A_{2,n}\exp (-ik_{n}r_{n}),
\end{eqnarray}
with $k_{j}=\sqrt{2\mu (Q-V_{j})/\hbar ^{2}}$, $r_{j}=r_{1}+(j-1)h$
and $V_{j}=[V(r_{j})+V(r_{j+1})]/2$. The wave function outside of
the barrier is given by
\begin{eqnarray}
u_{0} &=&A_{1,0}\exp (ik_{0}r_{0})+A_{2,0}\exp (-ik_{0}r_{0}),  \nonumber \\
u_{n+1} &=&A_{1,n+1}\exp (ik_{n+1}r_{n+1}),
\end{eqnarray}
with $k_{0}=k_{n+1}=\sqrt{2\mu Q/\hbar ^{2}}$ where $A_{1,0}$ and
$A_{1,n+1}$ are the amplitude of incident wave and transmitted wave,
respectively. By using the connection condition of wave function,
one can deduce the transmission amplitude and reflection amplitude
for the $n$th square barrier
\begin{eqnarray}
A_{1,n} &=&\frac{1}{2}\left( 1+\frac{k_{n+1}}{k_{n}}\right) \exp
\left[
i\left( k_{n+1}-k_{n}\right) r_{n}\right] A_{1,n+1}, \\
A_{2,n} &=&\frac{1}{2}\left( 1-\frac{k_{n+1}}{k_{n}}\right) \exp
\left[ i\left( k_{n+1}+k_{n}\right) r_{n}\right] A_{1,n+1},
\end{eqnarray}
and for the $j$th ($j<n$) square barrier
\begin{equation}
A_{1,j}=\frac{1}{2}\exp \left( -ik_{j}r_{j}\right) \left[ \exp
\left( ik_{j+1}r_{j}\right) (1+\frac{k_{j+1}}{k_{j}})A_{1,j+1}+\exp
\left( -ik_{j+1}r_{j}\right)
(1-\frac{k_{j+1}}{k_{j}})A_{2,j+1}\right],
\end{equation}%
\begin{equation}
A_{2,j}=\frac{1}{2}\exp \left( ik_{j}r_{j}\right) \left[ \exp \left(
ik_{j+1}r_{j}\right) (1-\frac{k_{j+1}}{k_{j}})A_{1,j+1}+\exp \left(
-ik_{j+1}r_{j}\right) (1+\frac{k_{j+1}}{k_{j}})A_{2,j+1}\right].
\end{equation}
The penetration probability is given by
\begin{equation}
P=\frac{\left\vert A_{1,n+1}\right\vert ^{2}}{\left\vert
A_{1,0}\right\vert ^{2}}.\label{tt}
\end{equation}
Normalization won't help-this is not a normalizable state. We choose
$A_{1,n+1}=1$, and then the $A_{1,0}$ can be recured according to
the above formulas and hence one can obtain the penetrability. As a
matter of fact, our method is a kind of numerical method to solve
one-dimensional Schr\"odinger equation for unbound state, in which
the differential equation is translated to recursion formulas.

Before we perform the calculation for $\alpha$, proton and cluster
radioactivity, we have checked this method with a soluble example
\begin{equation}
V(x)=V_{0}\cosh ^{-2}(x/a),V_{0}>0,
\end{equation}
for which the exact analytic transmission probability is known
\cite{LD,CE}. It is found that the result with this method
completely coincides with the analytic one, which confirms the
reliability of this method. Taking this method as a standard, one
can test whether the WKB approximation works well or not. In the WKB
approximation, the formula
\begin{equation}
P_{\text{WKB}}=\exp \left[ -\frac{2}{\hbar
}\int_{R_{\text{in}}}^{R_{\text{out}}}\sqrt{2\mu \left(
V(r)-Q\right) }dr\right],
\end{equation}
is employed to evaluate the penetrability, and one can estimate the
relative deviation $RD=(P_{\text{WKB}}-P)/P\times 100\%$ of this WKB
method. As a semiclassical approximation, there exist two classical
turning points $r_{\text{in}}$ and $r_{\text{out}}$ in WKB method.
The penetration only performs between $r_{\text{in}}$ and
$r_{\text{out}}$ and the effects of the regions I and III of
potential barrier in Figure 1(a) are neglected. However, according
to quantum mechanics, the particle can be also reflected back in the
region III with some probability. Additionally, the WKB method also
brings some errors when one evaluates the penetrability from
$r_{\text{in}}$ to $r_{\text{out}}$ (the region II) and it cannot
deal with these $Q$ values being near or larger than the top of the
barriers in principle while our fully quantum mechanical approach
has no such a drawback.

We select $\alpha$ decay events with $52 \leqslant Z\leqslant 118$,
and the experimental $Q$ values are taken from Refs.
\cite{ROYER,YS0}. For $\alpha$ decay, the $RD$ values have been
presented in Figure 2. The WKB approximation underestimates the
penetration probability by about $-40\%\sim -30\%$. It is not
possible to calculate the $\alpha$ decay half-life theoretically
with a high accuracy within the framework of barrier penetration
because the preformation factor is very difficult to be estimated
microscopically and the $\alpha$-daughter interaction has not yet
been well determined. From this point of view, the WKB approximation
thus works well in investigations of $\alpha$ decay especially for
SHN since the experimental half-lives of SHN tend to have a large
uncertainty. Indeed, because the deviation is nearly a constant in
the whole mass region, this constant error can be compensated by
other quantities such as a phenomenological assault frequency in
actual calculations within the WKB framework. In an analogous way,
we investigate the deviation of the WKB approximation for proton and
cluster radioactivity. The study of the nuclei far away from the
$\beta$-stable line has attracted world wide attention from both the
experimental and theoretical points of view. In the case of very
proton-rich nuclei, it is expected to observe the proton emission
experimentally \cite{DD1}. Since around 1980, the cluster
radioactivity was observed in experiments with daughter nuclei being
almost closed-shell spherical nuclei around $^{208}$Pb. The proton
and cluster emission can be treated as simple quantum tunneling
effects through a potential barrier just as the $\alpha$ decay
\cite{book,PRC}. We select cluster emitters with emitted particles
from $^{14}$C to $^{34}$Si and spherical proton emitters, for which
the experimental $Q$ values are taken from Refs. \cite{data,DD1,DE}.
Figure 3(a) shows the relative deviation $RD$ of the WKB method for
the proton radioactivity of spherical proton emitters. As can be
seen, the WKB method underestimates the penetrability by $-40\%\sim
-20\%$ and again the deviation does not fluctuate with a large
amplitude as that for $\alpha$ decay. Figure 3(b) presents the
relative deviation $RD$ of the WKB method for cluster radioactivity.
It indicates the WKB approximation works well for the cluster
emission with a deviation by only $-5\%\sim 15\%$. The $RD$ is found
to be insensitive to the nuclear potential $V_{N}(r)$ for these
three decay modes which indicates the conclusion we draw here is
universal.

\section{Summary}\label{sec5}\noindent

To summarize, the $\alpha$ decay half-lives of newly synthesized SHN
have been investigated in terms of the correlation between the
half-lives of $\alpha$ decay. The results of the present
calculations with this relationship based on the Royer's and VSS
formulas are in excellent agreement with the experimental data which
indicates the predictive power of our method. According to the
present calculations, an important conclusion is that the
experimental half-lives are themselves consistent with each other
confirming the reliability of the experimental observations and
measurements to a great extent. For the nuclei $^{282}$113 and
$^{279}$111, the half-lives from the theoretical estimations are
underestimated by one order of magnitude possibly due to nonzero
angular momentum transfer or some nuclear structure effects. The
half-lives of the synthesized SHN does not exhibit any evidence to
show that $Z=114$ and $N=172$ are shell closures in the considered
region according to our analysis. The other task of the present work
was to test the applicability of the WKB approximation for quantum
mechanical tunneling probabilities. We calculated the barrier
penetrability for $\alpha$ decay, proton and cluster emission
accurately with the recursion formulas by dividing the potential
barrier into a sequence of square barriers, which is a numerical
method to solve one-dimensional Schr\"odinger equation for unbound
state, and the results are compared with those of the WKB
approximation. It is found that the WKB method produces relative
deviations by about $-40\%\sim -30\%$ for $\alpha$ decay of heavy
and superheavy nuclei, $-40\%\sim -20\%$ for proton emission and
$-5\%\sim 15\%$ for cluster radioactivity. Also, in consideration of
the deviations being nearly constants in each decay mode, indeed,
the WKB approximation works well for these decays.

\section*{Acknowledgements}
Jianmin Dong is thankful to Prof. Jianzhong Gu and Hongfei Zhang for
providing useful help. This work is supported by the National
Natural Science Foundation of China (Grant Nos. 10875151, 10575119,
10975190 and 10947109), the Major State Basic Research Developing
Program of China under Grant Nos. 2007CB815003 and 2007CB815004, the
Knowledge Innovation Project (KJCX2-EW-N01) of Chinese Academy of
Sciences, CAS/SAFEA International Partnership Program for Creative
Research Teams (Grant No. CXTD-J2005-1), the Funds for Creative
Research Groups of China (Grant 11021504) and the financial support
from DFG of Germany.

\renewcommand{\baselinestretch}{1.0}

\newpage
%%%%%%%%%%%%%%%%%%%%%%%%%%%%%%%%%%%%%%%%%%%%%%%%%%%%%%%%%%%%%%%%%%%%%%%%%%%%%%%%%%%%%%%%%%%%%%%%%
\begin{table*}[h]
\label{table1} \caption{Calculated $\alpha$-decay half-lives of
recently synthesized SHN $^{A}Z$ within Eqs. (\ref{C}) and (\ref{D})
taking the nuclei in the isotope chains as references. The first
eight nuclei and the rest ones are obtained with the reference
nuclei $^{A-2}Z$ and $^{A+2}Z$, respectively.  The experimental data
\cite{YS0} and other theoretical results are also listed for
comparison. Some experimental $Q$ values are obtained by using the
measured $\alpha$ kinetic energies taking into account the electron
shielding corrections.}
\begin{tabular}{llllllll}
\hline
Nucleus& $Q^{\text{expt.}}$&  $T^{\text{expt.}}$&  $T^{(\ref{C})}$& $T^{(\ref{D})}$  & $ T^{\text{DDM3Y}}$ \cite{DDM3Y2}& $T^{\text{GLDM}}$ \cite{Royer0,GLDM3}  \\
%Nuclei&$Q$[MeV]&$Q$[MeV]&$T_{1/2}$&$T_{1/2}(Q_{ex})$&$T_{1/2}$&$T_{1/2}(Q_{Audi})$&$T_{1/2}(Q_{ex})$&$T_{1/2}(Q_{Audi})$\\
\hline
 $^{293}$116&$10.67(6)$& $53^{+62}_{-19}$ ms &$62.0^{+75.8}_{-20.7}$ ms &$66.1^{+80.8}_{-22.0}$ ms &$206^{+90}_{-61}$ ms &$22.81^{+10.22}_{-7.06}$ ms       \\
%&&&&&&& \\
$^{292}$116&$10.80(7)$& $18^{+16}_{-6}$ ms  &$21.2^{+9.5}_{-5.1}$ ms &$22.7^{+10.2}_{-5.4}$ ms &$39^{+20}_{-13}$ ms  &$10.45^{+5.65}_{-3.45}$ ms        \\
%&&&&&&& \\
$^{289}$115&$10.50(9)$& $0.22^{+0.26}_{-0.08}$ s  &$0.13^{+0.61}_{-0.06}$ s &$0.13^{+0.65}_{-0.06}$ s &--  &--        \\
%&&&&&&& \\
$^{289}$114& $9.96(6)$& $2.7^{+1.4}_{-0.7}$ s &$1.6^{+0.54}_{-0.31}$ s &$1.7^{+0.58}_{-0.33}$ s &$3.8^{+1.8}_{-1.2}$ s &$0.52^{+0.25}_{-0.17}$ s    \\
%&&&&&&& \\
$^{284}$113  &$10.15(6)$& 0.48$^{+0.58}_{-0.17}$ s &$5.6^{+10.3}_{-2.2}$ s &$4.3^{+7.9}_{-1.7}$ s &1.55$^{+0.72}_{-0.48}$ s&0.43$^{+0.21}_{-0.13}$ s       \\
%&&&&&&& \\
$^{285}$112&$9.29(6)$& $34^{+17}_{-9}$ s&    $50.2^{+15.9}_{-9.3}$ s & $52.5^{+16.6}_{-9.7}$ s &$75^{+41}_{-26}$ s &$13.22^{+7.25}_{-4.64}$ s  \\
%&&&&&&& \\
$^{280}$111  & $9.87(6)$&  3.6 $^{+4.3}_{-1.3}$  s  &$2.9^{+5.2}_{-1.2}$ s &$1.9^{+3.4}_{-0.8}$ s&1.9$^{+0.9}_{-0.6}$ s  &0.69$^{+0.33}_{-0.23}$ s     \\
%&&&&&&& \\
  $^{291}$116&$10.89(7)$&$18^{+22}_{-6}$ ms &$15.4^{+18.0}_{-5.5}$ ms &$14.4^{+16.9}_{-5.2}$ ms &$60.4^{+30.2}_{-20.1}$ ms &$6.35^{+3.15}_{-2.08}$ ms     \\
%&&&&&&& \\
 $^{290}$116&$11.00(8)$& $7.1^{+3.2}_{-1.7}$ ms &$6.0^{+5.4}_{-2.0}$ ms &$5.6^{+5.0}_{-1.9}$ ms  &$13.4^{+7.7}_{-5.2}$ ms &$3.47^{+1.99}_{-1.26}$ ms  \\
%&&&&&&& \\
  $^{287}$115  & $10.74(9)$ &   32$^{+155}_{-14}$ ms & $55.5^{+65.5}_{-20.2}$ ms & $52.2^{+61.7}_{-19.0}$ ms  &   $51.7^{+35.8}_{-22.2}$  ms   &46.0$^{+33.1}_{-19.1}$ ms \\
%&&&&&&& \\
  $^{287}$114&$10.16(6)$& $0.48^{+0.16}_{-0.09}$ s& $0.79^{+0.41}_{-0.21}$ s & $0.74^{+0.39}_{-0.19}$ s &$1.13^{+0.52}_{-0.40}$ s  &$0.16^{+0.08}_{-0.05}$ s  \\
%&&&&&&& \\
$^{282}$113&$10.83(8)$& $73^{+134}_{-29}$ ms &$6.3^{+7.6}_{-2.2}$ ms&  $8.1^{+9.8}_{-2.9}$ ms &--&$7.8^{+4.6}_{-2.8}$ ms  \\
%&&&&&&& \\
  $^{283}$112& $9.67(6)$& $3.8^{+1.2}_{-0.7}$ s &$2.6^{+1.3}_{-0.7}$ s &$2.5^{+1.2}_{-0.7}$ s &$5.9^{+2.9}_{-2.0}$ s &$0.95^{+0.48}_{-0.32}$ s    \\
%&&&&&&& \\
$^{278}$111&$10.89(8)$& $4.2^{+7.5}_{-1.7}$ ms&$5.2^{+6.2}_{-1.9}$ ms &$8.0^{+9.6}_{-2.9}$ ms &-- &$1.5^{+0.9}_{-0.5}$ ms \\
%&&&&&&& \\
\hline
\end{tabular}
\end{table*}
%%%%%%%%%%%%%%%%%%%%%%%%%%%%%%%%%%%%%%%%%%%%%%%%%%%%%%%%%%%%%%%%%%%
%%%%%%%%%%%%%%%%%%%%%%%%%%%%%%%%%%%%%%%%%%%%%%%%%%%%%%%%%%%%%%%%%%%%%%%%%%%%%%%%%%%%%%%%%%%%

\newpage
%%%%%%%%%%%%%%%%%%%%%%%%%%%%%%%%%%%%%%%%%%%%%%%%%%%%%%%%%%%%%%%%%%%%%%%%%%%%%%%%%%%%%%%%%%%%%%%%
\begin{table*}[h]
\label{table2} \caption{Calculated $\alpha$-decay half-lives of
recently synthesized SHN $^{A}Z$ within Eqs. (\ref{C}) and (\ref{D})
taking the nuclei in $\alpha$-decay chains as references. The
results in columns 4 and 5 are obtained with the help of their
daughter nuclei $^{A-4}(Z-2)$ while the columns 6 and 7 are the
estimated ones with the help of their parent nuclei $^{A+4}(Z+2)$.
The experimental data \cite{YS0} and other theoretical results are
also listed for comparison. Some experimental $Q$ values are
obtained using the measured $\alpha$ kinetic energies taking into
account the electron shielding corrections.}
\begin{tabular}{llllllllllllllllllll}
\hline
Nucleus& $Q^{\text{expt.}}$&  $T^{\text{expt.}}$&  $T^{(\ref{C})}$ &  $T^{(\ref{D})}$&  $T^{(\ref{C})}$ &  $T^{(\ref{D})}$& $ T^{\text{DDM3Y}}$ \cite{DDM3Y2}& $T^{\text{GLDM}}$ \cite{Royer0,GLDM3}  \\
%Nuclei&$Q$[MeV]&$Q$[MeV]&$T_{1/2}$&$T_{1/2}(Q_{ex})$&$T_{1/2}$&$T_{1/2}(Q_{Audi})$&$T_{1/2}(Q_{ex})$&$T_{1/2}(Q_{Audi})$\\
\hline
  $^{294}$118&$11.81(6)$& $0.89^{+1.07}_{-0.31}$ ms &$0.31^{+0.14}_{-0.07}$ ms &$0.32^{+0.14}_{-0.08}$ ms& -- & -- &$0.66^{+0.23}_{-0.18}$ ms &$0.15^{+0.05}_{-0.04}$ ms   \\
%&&&&&&& \\
 $^{290}$116&$11.00(8)$& $7.1^{+3.2}_{-1.7}$ ms &--   &-- &$20.5^{+24.7}_{-7.2}$ ms &$19.8^{+23.8}_{-6.9}$ ms &$13.4^{+7.7}_{-5.2}$ ms &$3.47^{+1.99}_{-1.26}$ ms  \\
%&&&&&&& \\
 $^{293}$116&$10.67(6)$& $53^{+62}_{-19}$ ms &$136^{+71}_{-35}$ ms &$136^{+70}_{-35}$ ms&--&--&$206^{+90}_{-61}$ ms &$22.81^{+10.22}_{-7.06}$ ms       \\
%&&&&&&& \\
  $^{289}$114& $9.96(6)$& $2.7^{+1.4}_{-0.7}$ s &$1.6^{+0.8}_{-0.4}$ s&$1.6^{+0.8}_{-0.4}$ s &$1.0^{+1.2}_{-0.4}$ s&$1.1^{+1.2}_{-0.4}$ s   &$3.8^{+1.8}_{-1.2}$ s  &$0.52^{+0.25}_{-0.17}$ s    \\
%&&&&&&& \\
   $^{285}$112&$9.29(6)$& $34^{+17}_{-9}$ s&--&--&   $56.3^{+29.2}_{-14.6}$ s& $55.9^{+29.0}_{-14.5}$ s  &$75^{+41}_{-26}$ s &$13.22^{+7.25}_{-4.64}$ s  \\
%&&&&&&& \\
  $^{292}$116&$10.80(7)$& $18^{+16}_{-6}$ ms  &$40.9^{+16.4}_{-9.2}$ ms&$43.1^{+17.2}_{-9.7}$ ms&--&--&$39^{+20}_{-13}$ ms  &$10.45^{+5.65}_{-3.45}$ ms        \\
%&&&&&&& \\
  $^{288}$114&$10.09(7)$& $0.8^{+0.32}_{-0.18}$ s &--&--& $0.35^{+0.31}_{-0.12}$ s& $0.33^{+0.30}_{-0.11}$ s &$0.67^{+0.37}_{-0.27}$ s &$0.22^{+0.12}_{-0.08}$ s   \\
%&&&&&&& \\
  $^{291}$116&$10.89(7)$&$18^{+22}_{-6}$ ms &$23.9^{+8.0}_{-4.5}$ ms &$23.9^{+8.0}_{-4.5}$ ms&--&--&$60.4^{+30.2}_{-20.1}$ ms &$6.35^{+3.15}_{-2.08}$ ms     \\
%&&&&&&& \\
  $^{287}$114&$10.16(6)$& $0.48^{+0.16}_{-0.09}$ s& $0.71^{+0.22}_{-0.13}$ s& $0.70^{+0.22}_{-0.13}$ s & $0.36^{+0.44}_{-0.12}$ s& $0.36^{+0.44}_{-0.12}$ s &$1.13^{+0.52}_{-0.40}$ s  &$0.16^{+0.08}_{-0.05}$ s  \\
%&&&&&&& \\
  $^{283}$112& $9.67(6)$& $3.8^{+1.2}_{-0.7}$ s &--  &--   &$2.6^{+0.9}_{-0.5}$ s&$2.6^{+0.9}_{-0.5}$ s &$5.9^{+2.9}_{-2.0}$ s &$0.95^{+0.48}_{-0.32}$ s    \\
%&&&&&&& \\
  $^{288}$115  &$10.61(6)$&87 $^{+105}_{-30}$ ms & $116^{+140}_{-41}$ ms & $121^{+146}_{-43}$ ms &--&--  & $410.5^{+179.4}_{-122.7}$ ms  &94.7$^{+41.9}_{-28.9}$ ms   \\
%&&&&&&& \\
  $^{284}$113  &$10.15(6)$& 0.48$^{+0.58}_{-0.17}$ s &$2.8^{+3.3}_{-1.0}$ s&$2.7^{+3.2}_{-1.0}$ s &$0.36^{+0.43}_{-0.12}$ s&$0.34^{+0.42}_{-0.12}$ s  &1.55$^{+0.72}_{-0.48}$ s&0.43$^{+0.21}_{-0.13}$ s       \\
%&&&&&&& \\
  $^{280}$111  & $9.87(6)$&  3.6 $^{+4.3}_{-1.3}$  s  &$3.4^{+4.1}_{-1.2}$ s &$2.9^{+3.5}_{-1.0}$ s  &$0.61^{+0.74}_{-0.22}$ s &$0.64^{+0.78}_{-0.23}$ s  &1.9$^{+0.9}_{-0.6}$ s  &0.69$^{+0.33}_{-0.23}$ s     \\
%&&&&&&& \\
  $^{276}$109  & $9.85(6)$& 0.72$^{+0.87}_{-0.25}$ s & $0.36^{+0.43}_{-0.13}$ s  & $0.45^{+0.53}_{-0.16}$ s   & $0.75^{+0.90}_{-0.27}$ s  & $0.90^{1.08}_{-0.33}$ s &0.45$^{+0.23}_{-0.14}$ s&0.19$^{+0.08}_{-0.06}$ s      \\
%&&&&&&& \\
  $^{272}$107  & $9.15(6)$&  9.8$^{+11.7}_{-3.5}$  s &--&--& $19.6^{+23.7}_{-6.8}$ s & $15.8^{+19.1}_{-5.5}$ s  &10.1$^{+5.4}_{-3.4}$ s&5.12$^{+3.19}_{-1.58}$  s     \\
  %&&&&&&& \\
  $^{287}$115  & $10.74(9)$ &   32$^{+155}_{-14}$ ms & $22.6^{+110.6}_{-10.2}$ ms & $23.2^{+113.5}_{-10.4}$ ms &--&-- &   $51.7^{+35.8}_{-22.2}$  ms   &46.0$^{+33.1}_{-19.1}$ ms \\
%&&&&&&& \\
  $^{283}$113  & $10.26(9)$&100$^{+490}_{-45}$ ms  & $3.6^{+17.3}_{-1.7}$ s & $3.5^{+16.5}_{-1.6}$ s & $141^{+684}_{-62}$ ms & $138^{+669}_{-60}$ ms &201.6$^{+164.9}_{-84.7}$ms&222$^{+172}_{-96}$ ms     \\
%&&&&&&& \\
  $^{279}$111  & $10.52(16)$&170$^{+810}_{-80}$ ms &$33^{+155}_{-15}$ ms &$32^{+153}_{-15}$ ms &$4.7^{+22.8}_{-2.1}$ ms &$4.9^{+24.0}_{-2.2}$ ms  &9.6$^{+14.8}_{-5.7}$ ms&12.4$^{+19.9}_{-7.6}$ ms      \\
  %&&&&&&& \
 $^{275}$109  & $10.48(9)$& 9.7$^{+46}_{-4.4}$  ms & --&--& $50.3^{+239.6}_{-23.7}$ ms&$51.0^{+243.0}_{-24.0}$ ms &2.75$^{+1.85}_{-1.09}$ ms&4.0$^{+2.8}_{-1.6}$ ms       \\
 %&&&&&&& \\
$^{282}$113&$10.83(8)$& $73^{+134}_{-29}$ ms &$29.6^{+52.8}_{-12.0}$ ms &$23.8^{+42.6}_{-9.6}$ ms&--&--&--&$7.8^{+4.6}_{-2.8}$ ms  \\
%&&&&&&& \\
$^{278}$111&$10.89(8)$& $4.2^{+7.5}_{-1.7}$ ms&$5.5^{+10.1}_{-2.1}$ ms &$7.4^{+13.5}_{-2.8}$ ms &$10.3^{+18.9}_{-4.1}$ ms &$12.9^{+23.6}_{-5.1}$ ms&-- &$1.5^{+0.9}_{-0.5}$ ms \\
%&&&&&&& \\
$^{274}$109&$9.95(10)$& $440^{+810}_{-170}$ ms &$832^{+3982}_{-382}$ ms &$1122^{+5370}_{-515}$ ms&$335^{+598}_{-135}$ ms &$251^{+449}_{-102}$ ms&--&$108^{+96}_{-51}$ ms   \\
%&&&&&&& \\
$^{270}$107&$9.11(8)$& $61^{+292}_{-28}$ s &--&--&$32.1^{+59.1}_{-12.4}$ s &$23.9^{+44.0}_{-9.2}$ s&--&$7.7^{+6.1}_{-3.3}$ s  \\
%&&&&&&& \\
\hline
\end{tabular}
\end{table*}
%%%%%%%%%%%%%%%%%%%%%%%%%%%%%%%%%%%%%%%%%%%%%%%%%%%%%%%%%%%%%%%%%%%%%%%%%%%%%%%%%%%%%%%%%%%%

\newpage
%%%%%%%%%%%%%%%%%%%%%%%%%%%%%%%%%%%%%%%%%%%%%%%%%%%%%%%%
\begin{table}[h]
\label{table3} \caption{Predicted $\alpha$-decay half-lives of the
nuclei $^{A}Z$ in $\alpha$-decay chain starting from $^{294}$117
within Eqs. (\ref{C}) and (\ref{D}). The experimental $Q$ values are
obtained with the measured $\alpha$ kinetic energies \cite{117}
taking account of the electron shielding corrections. The results in
columns 3 and 4 are obtained with their isotopes $^{A-2}Z$ and the
ones in columns 5 and 6 are obtained with the help of $^{294}$117.}
\begin{tabular}{llllllllllllllll}
\hline nuclei& $Q^{\text{Exp.}}$(MeV)
&$T^{(\ref{C})}$ &$T^{(\ref{D})}$ &$T^{(\ref{C})}$ &$T^{(\ref{D})}$  \\
\hline
%&&&&&&& \\
$^{290}$115& 10.14(41) & $1.9^{+2.3}_{-0.7}$ s & $1.6^{+1.9}_{-0.6}$ s   & $4.8^{+22.6}_{-2.2}$ s&$3.7^{+17.7}_{-1.7}$ s  \\
$^{286}$113& 9.81(10)  &$4.8^{+5.8}_{-1.7}$ s  &$4.3^{+5.2}_{-1.5}$ s&$9.0^{+42.7}_{-4.2}$ s  &$7.2^{+34.1}_{-3.3}$ s    \\
$^{282}$111& 9.18(10)  &$9.0^{+10.8}_{-3.3}$ min &$6.6^{+7.8}_{-2.4}$ min &$2.9^{+13.8}_{-1.3}$ min&$2.0^{+9.3}_{-0.9}$ min  \\
$^{278}$109& 9.74(19)  &$1.4^{+1.7}_{-0.5}$ s  &$1.5^{+1.8}_{-0.5}$ s &$0.48^{+2.3}_{-0.2}$ s  &$0.54^{+2.6}_{-0.3}$ s  \\
$^{274}$107& 8.98(10)  &$33.2^{+39.6}_{-11.9}$ s  &$32.5^{+38.8}_{-11.6}$ s  &$22.4^{+106.5}_{-10.4}$ s  &$19.5^{+92.5}_{-9.0}$ s    \\
\hline
%&&&&&&& \\
\end{tabular}
\end{table}
%%%%%%%%%%%%%%%%%%%%%%%%%%%%%%%%%%%%%%%%%%%%%%%%%%%%%%%%%%%%

\newpage
% Figure 1
\begin{figure}[htb]
  \begin{center}
  \includegraphics[width=0.9\textwidth]{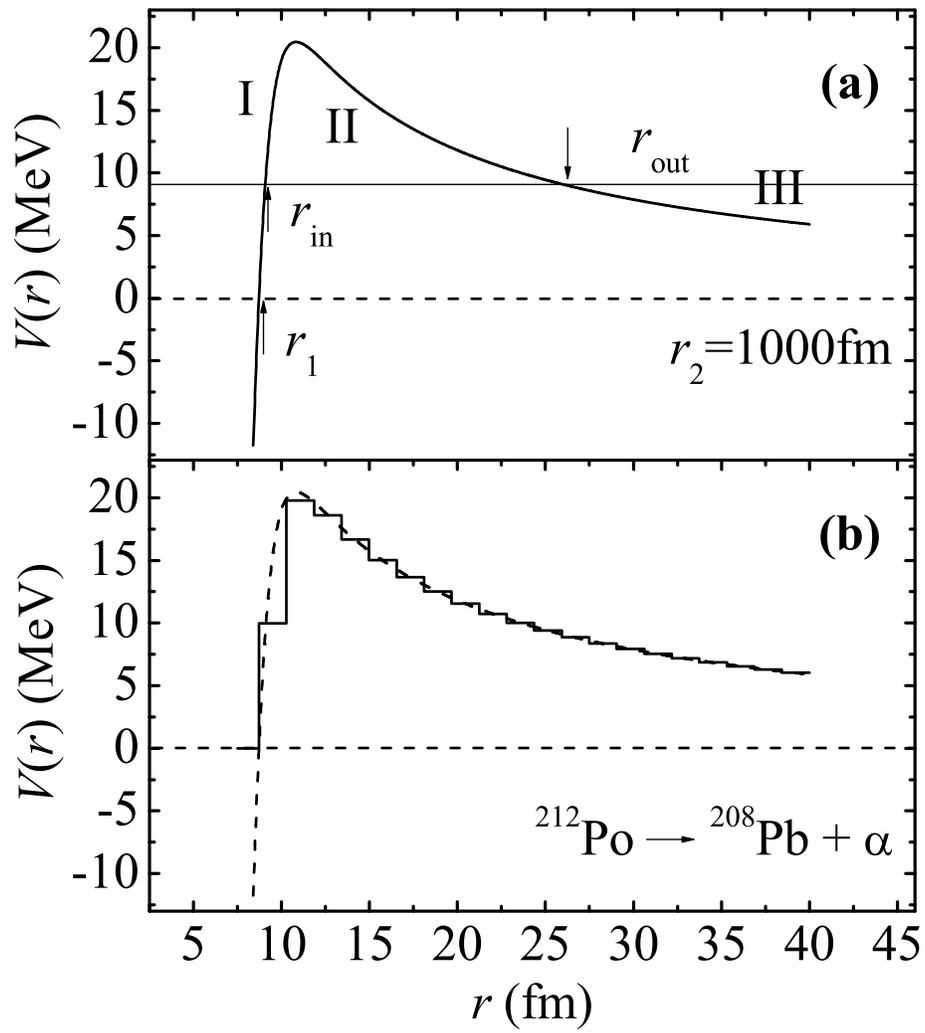}
  \end{center}
  \caption{(a) Potential barrier of $^{212}$Po$\rightarrow
 ^{208}$Pb$+\alpha$. (b) It is divided into a sequence of square barriers.}
\end{figure}

\newpage
% Figure 2
\begin{figure}[htb]
  \begin{center}
  \includegraphics[width=0.9\textwidth]{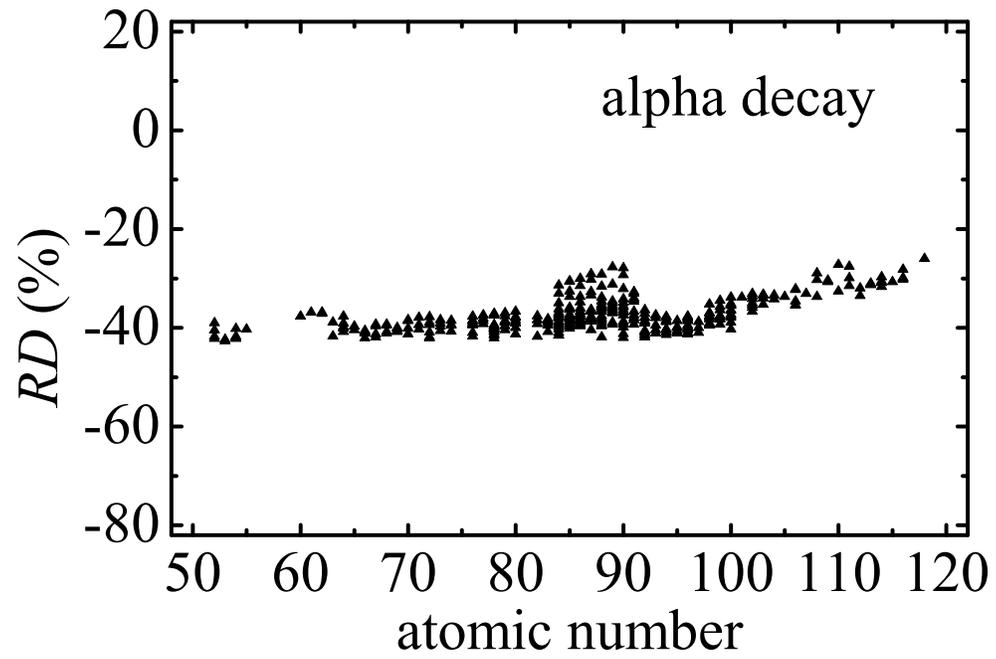}
  \end{center}
  \caption{Relative deviation of penetrability caused by the WKB approximation for $\alpha$ decay.}
  \end{figure}

\newpage
% Figure 3
\begin{figure}[htb]
  \begin{center}
  \includegraphics[width=0.9\textwidth]{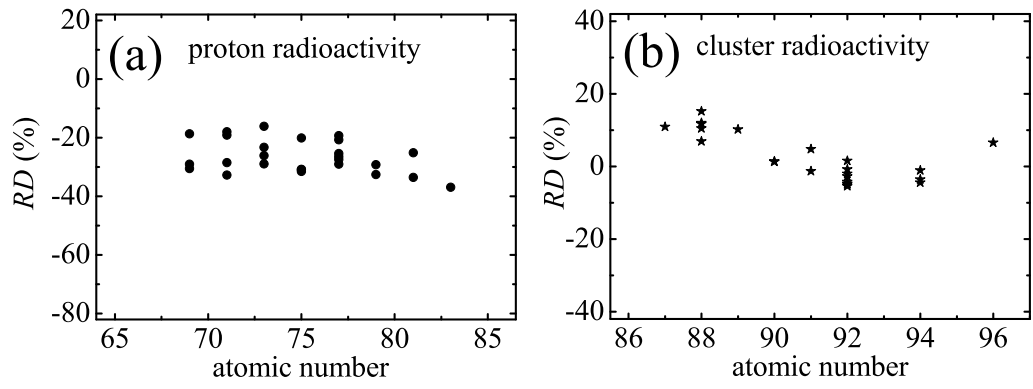}
  \end{center}
  \caption{Relative deviation of penetrability caused by the WKB approximation for proton and cluster radioactivity.}
\end{figure}

\bigskip
\end{document}